\documentclass[aps,prx,reprint]{revtex4-2}

\usepackage[utf8]{inputenc}
\usepackage{microtype}
\usepackage{xspace}
\usepackage{mathrsfs}
\usepackage{amsfonts,mathtools,mathrsfs}
\usepackage{bm}
\usepackage{xcolor}
\usepackage[colorlinks=true, linkcolor = blue, hyperfootnotes=false, citecolor = blue]{hyperref}
\newcommand{\R}{\mathbb{R}}

\newcommand{\tp}{\otimes}
\newcommand{\str}{\varepsilon}
\newcommand{\norm}[1]{\left\lVert #1 \right\rVert}

\newcommand{\scalar}[1]{\left\langle{#1}\right\rangle}

\renewcommand{\t}{\mathbf}
\newcommand{\gt}{\bm}
\renewcommand\[{\begin{equation}}
\renewcommand\]{\end{equation}}
\let\div\relax
\DeclareMathOperator{\div}{div}

\DeclareMathOperator{\cof}{cof}
\newcommand{\comment}[1]{#1}

\begin{document}
\title{A topological counting rule for shells}
\author{Hussein Nassar}\email{Corresponding author: hnassar@uh.edu}
\affiliation{Department of Mechanical and Aerospace Engineering, University of Houston, Houston TX 77204}

\begin{abstract}
Holding a shell in their hands, one can apply six loads: three by pulling and shearing, and three by bending and twisting. Here, it is shown that the shell resists exactly three load cases and comply with the other three, provided the shell is simply connected, meaning it has no holes and no handles. Formally, it is shown that the space of homogeneous membrane and bending strains, defined in the sense of plate theory, that can be relaxed into an infinitesimal isometry by a periodic, or a statistically homogeneous, deflection is three-dimensional for any simply-connected periodic, or statistically homogeneous, shell, be it corrugated, creased or wrinkled.
\end{abstract}
\maketitle

Maxwell~\cite{Maxwell1864} famously derived a counting rule for the zero modes of a truss. These are motions of the truss that do not stretch any of its members. The rule assigns one zero mode for each degree of freedom in excess of the number of drawn elastic bonds. Calladine~\cite{Calladine1978a, Pellegrino1986} refined the rule to account for redundant bonds: adding a bond either reduces indeterminacy by eliminating a zero mode or increases redundancy by creating a state of self-stress. The latter is a field of bar tensions in equilibrium with no external forces or reactions. Equivalently, a state of self-stress represents a way to redistribute tension from one bar to others, hence a redundancy.

For isostatic trusses, meaning with as many degrees of freedom as bonds, Kane and Lubensky~\cite{Kane2013} proposed a local rule that counts the zero modes supported by a subsystem. Unlike Maxwell–Calladine, theirs is sensitive to the mode shape, particularly to where and how the mode localizes along the subsystem’s boundary. Such sensitivity can only arise in sufficiently large trusses, where localization effects can manifest. Their rule also explains the emergence of robust, localized zero modes, analogous to protected electronic boundary states, using the topology of the spectrum of the compatibility matrix rather than the topology of the truss itself. Consequently, the Kane–Lubensky rule has powerful implications for the behavior of free surfaces in isostatic lattice materials, such as during indentation~\cite{Bilal2017}, crack initiation, or crack growth~\cite{deWaal2025}.

In cases where the truss triangulates a shell, its zero modes are discrete counterparts of the shell’s isometric deformations. These deformations, a classical subject in surface geometry, have regained prominence with advances in space deployable structures~\cite{Haraszti2025}, soft robotics~\cite{Yang2021}, 4D printing~\cite{Holmes2011} and soft matter physics more broadly~\cite{Tobasco2022, Plummer2020}, where they are used to geometrically encode or model morphing paths and buckled shapes. Conversely, the absence of such deformations is indicative of a structurally efficient shell, one that carries loads by surface tension, needless of bending stresses~\cite{ciarlet2006}. Beyond what is directly implied by Maxwell-Calladine, results reminiscent of counting rules for surfaces are rare. Notable ones include ($i$) rigidity results (e.g., for convex polyhedra~\cite{Connelly1993}), where the count is zero; ($ii$) Cohn-Vossen’s counting argument for surfaces of revolution~\cite{Audoly2010}; ($iii$) the uniqueness of the mechanism of a nondegenerate mesh of planar quads~\cite{Schief2008}; and ($iv$) Tachi’s conjecture on the existence of a two-degree-of-freedom family of periodic mechanisms in generic periodic triangular meshes~\cite{Tachi2015}. Other results provide local counts, e.g., for a single vertex or a neighborhood~\cite{Ivanova-Karatopraklieva1994,Ivanova-Karatopraklieva1995}, or derive isometric deformations for specific surfaces, e.g., origami tessellations and surfaces of translation \cite{Demaine2015a,Lang2018,Mundilova2019,Nassar2023,Karami2024a}.

\begin{figure*}
    \centering
    \includegraphics[width=\linewidth]{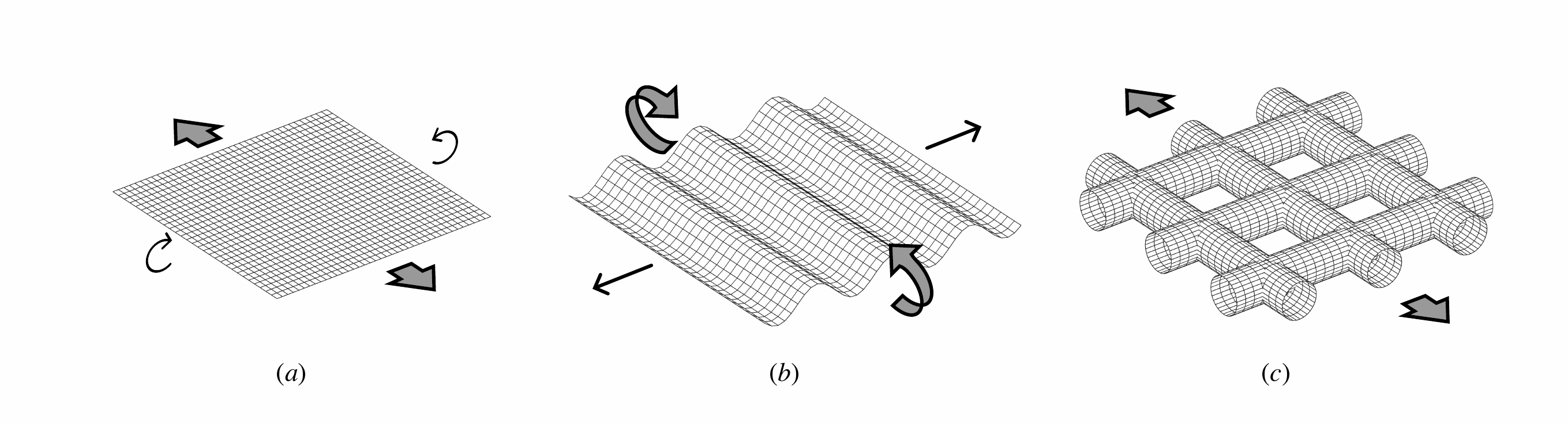}
    \caption{The static-geometric analogy for effective fields. Wide arrows are admissible stresses; thin arrows are isometric deformations: (a) in a plane, admissible longitudinal tension corresponds to isometric lateral bending; (b) in a singly corrugated membrane, an admissible longitudinal moment corresponds to isometric lateral extension; (c) in a multiply connected grid of tubes, admissible longitudinal tension has no corresponding isometric lateral bending.}
    \label{fig:analogy}
\end{figure*}

Here, a new counting rule is established for simply connected shells and trusses that discretize them. These trusses are locally isostatic but topologically trivial in the Kane–Lubensky sense. Nonetheless, the simply connected topology enables a count of effective isometric deformations, i.e., membrane-like or flexure-like deformations that are uniform up to a certain correction. The count is simple: there exist exactly three such isometric deformations. Conversely, a shell resists exactly three effective loads, tension and bending combined, provided it is simply connected, irrespective of its composition or geometry. Note that for modes to be meaningfully classified as membrane, flexure, or effective, the truss must be periodic or statistically homogeneous. The proof relies on two familiar results, seldom used in tandem: the static-geometric analogy, and the Hill-Mandel lemma. Both are introduced under the assumptions of smooth periodic graphs along with a proof of the counting rule. A discussion of the role of topology follows and the validity of the proof for more general, piecewise-smooth and/or non-periodic, shells is discussed.

\subsection*{The static-geometric analogy}
Consider a shell \comment{whose midsurface is the graph of a function~$z=z(x_1,x_2)$. A statically admissible distribution of membrane stresses $\gt\sigma$ satisfies two force balances, namely
\[\label{eq:eq}
    \div\gt\sigma = 0,\quad \t H_z:\gt\sigma = 0,
\]
where $\t H_z$ is the Hessian of $z$ \comment{and the colon denotes a double contraction}. The first is force balance of surface tensions (i.e., in the plane tangent to the shell); the second is force balance in the normal direction, identical to the well-known Laplace law (minus the internal pressure).} Letting $\gt\sigma=\cof \t H_\phi$ derive from an Airy stress function $\phi$ eliminates the first equation and recasts the second into Pucher's equation~\cite{Adiels2025}
\[\label{eq:HzHphi}
    \t H_z:\cof \t H_\phi = 0.
\]
Now if a displacement $(\t u\equiv(u_1,u_2), w)$ is isometric, then the infinitesimal membrane strain
\[\label{eq:comp}
\gt\str \equiv \gt\nabla^s\t u + \gt\nabla w\tp^s\gt\nabla z
\]
vanishes meaning that $-\gt\nabla w\tp^s\gt\nabla z=\gt\nabla^s\t u$ is a compatible planar strain. \comment{Therein, $\gt\nabla$ denote the gradient, $\tp$ is tensor product and the superscript $s$ denotes their symmetric variant.} The compatibility equation, expanded and reduced, turns out to be identical to~\eqref{eq:HzHphi} with $w$ replacing~$\phi$. This is the essence of the static-geometric analogy: a deflection is isometric if, as a stress function, it is statically admissible, and conversely.

The analogy is well-known. \comment{Calladine in particular noted that $\t H_z:\cof\t H_\psi = 0$ describes the absence of transverse loads when the test field $\psi$ is an Airy stress function, and the conservation of Gauss curvature when the test field $\psi$ is a transverse deflection~\cite{Calladine1977}.} Here, the novelty shall be in completing the analogy with integral formulas for average strains and stresses. In particular, for a periodic shell, let
\[\label{eq:uw}
\begin{bmatrix}
    u_\alpha \\ w
\end{bmatrix} =
\begin{bmatrix}
    E_{\alpha\beta}x_\beta + z\chi_{\alpha\beta}x_\beta \\ -\frac{1}{2}\chi_{\mu\nu}x_\mu x_\nu
\end{bmatrix}
+ \text{periodic corrections}
\]
be the displacement field due to an effective membrane strain $\t E$ and an effective bending strain $\gt\chi$. This is but the displacement that would reign in a homogeneous plate of uniform thickness, up to a correction allowing to relax some of the strain energy in function of the shell's profile and composition. For a brief overview of this ansatz and of the elasticity of thin periodic plates, see~\cite{Sab2020}. The effective strains $(\t E,\gt\chi)$ can be recovered from the local strain measures $(\gt\str,\t H_w)$ by averaging. If the deformation is isometric, the averages $\scalar{\cdot}$ can be written as
\[\label{eq:EChi}
\t E_o = \scalar{z\t H_w}, \quad \gt\chi_o = -\scalar{\t H_w},
\]
thanks to \comment{ansatz~\eqref{eq:uw}}, the divergence theorem and the vanishing of boundary terms by periodicity. Unless otherwise indicated, the subscript~``$o$'' will be used to denote isometric strains and will be dropped when $\t E$ and $\gt\chi$ are meant to be general strains. Similarly, for a statically admissible periodic membrane stress $\gt\sigma$ of Airy stress $\phi$, the effective membrane and bending stresses are given by the resultants
\[\label{eq:resultant}
\gt\Sigma = \scalar{\gt\sigma} = \scalar{\cof \t H_\phi}, \quad \t M = \scalar{z\gt\sigma} = \scalar{z\cof \t H_\phi}.
\]
Thus, by comparing~\eqref{eq:EChi} and~\eqref{eq:resultant}, it comes that the static-geometric analogy further matches $-\cof\gt\chi_o$ and $\gt\Sigma$ as well as $\cof\t E_o$ and $\t M$. Overall, the analogy provides an isomorphism
\[\label{eq:iso}
\begin{bmatrix}
    \gt\Sigma \\ \t M
\end{bmatrix} =
\begin{bmatrix}
    \t 0 & -\cof \\ \cof & \t 0
\end{bmatrix}
\begin{bmatrix}
    \t E_o \\ \gt\chi_o
\end{bmatrix}
\]
between the spaces of effective isometric deformations and of effective admissible stresses, where the $\cof$ operator acts as a $90^\circ$ rotation; see Figure~\ref{fig:analogy}. \comment{Note that isometries always refer to there being no stretching of the shell's midsurface, i.e., to the vanishing of local strains~$\gt\str$. In particular, an effective deformation $(\t E,\gt\chi)$ is isometric if it is the outcome of a deflection with vanishing local strains.}


\subsection*{The counting rule}
What is lacking here, to apply arguments à la Maxwell-Calladine~\cite{Calladine1978a}, is to identify the spaces of effective admissible stresses and isometric strains with the image and kernel of some structural matrix. In the present context, the appropriate matrix turns out to be the matrix of effective elasticity moduli $\t A$. \comment{To define it, consider the minimization of the average membrane strain energy density $\scalar{\gt\str:\t c:\gt\str}$ over the periodic corrections of ansatz~\eqref{eq:uw} for prescribed $(\t E,\gt\chi)$. Therein, $\t c$ is the shell's membrane stiffness tensor. This is a partial minimization of a positive semi-definite quadratic form and results in another positive semi-definite quadratic form
\[\label{eq:min}
     \begin{bmatrix}
        \t E \\ \gt\chi
    \end{bmatrix}^T \t A \begin{bmatrix}
        \t E \\ \gt\chi
    \end{bmatrix}
    \equiv
    \min_{\text{ansatz \eqref{eq:uw}}} \scalar{\gt\str:\t c:\gt\str}
\]
whose coefficients are identified as $\t A$. The membrane strain energy conveniently ignores bending so that isometric deformations are zero modes of $\t A$}. In other words, for the purposes of counting, the shell is modeled as a membrane locally, and as a plate (Kirchhoff-Love) globally.

\comment{To conclude, the symmetry of $\t A$ provides an orthogonal decomposition
\[
    \text{Im}\t A \oplus \text{Ker}\t A = S^2\times S^2
\]
of the space of pairs of symmetric second order tensors $S^2\times S^2$.} Then, the \emph{claim}
\[\label{eq:imker}
    \text{Im}\t A = \{(\gt\Sigma,\t M)\}, \quad
    \text{Ker}\t A = \{(\t E_o,\gt\chi_o)\}
\]
recasts the same decomposition into
\[
    \{(\t\Sigma,\t M)\} \oplus \{(\t E_o,\gt\chi_o)\} = S^2\times S^2.
\]
The analogy, on the other hand, showed that these spaces are isomorphic:
\[
\{(\t\Sigma,\t M)\} \cong \{(\t E_o,\gt\chi_o)\}.
\]
Taking the dimensions then yields the counting rule
\[\label{eq:count}
    \dim \{(\t E_o,\gt\chi_o)\} = \frac12\dim(S^2\times S^2) = 3.
\]

\subsection*{The Hill-Mandel lemma}
However intuitive, claim~\eqref{eq:imker} remains heuristic and warrants some attention. For instance, it is not quite obvious that admissible effective stresses $(\gt\Sigma,\t M)$ can all be realized as the response of the shell to some imposed deflection for these responses could reasonably depend, say, on the composition of the shell. This turns out not to be the case thanks to a profound general result of homogenization theory known as the Hill-Mandel lemma~\cite{Hill1963}. The lemma states that the virtual stress-strain work is the same if computed using the local fields or the effective fields. Namely, for any admissible periodic $\gt\sigma$ and any admissible $(\t u,w)$ in the sense of~\eqref{eq:uw}, one has
\[\label{eq:HM}
    \scalar{\gt\sigma:\gt\str} = \gt\Sigma:\t E + \t M:\gt\chi.
\]
The proof is straightforward and employs the divergence theorem in conjunction with periodicity~\eqref{eq:uw}, admissibility~\eqref{eq:eq} and the compatibility conditions~\eqref{eq:comp}; the Supplemental Material contains a proof. Most importantly, the lemma ensures that the static definition of stresses, by the resultants~\eqref{eq:resultant}, is consistent with the thermodynamic definition
\[
\gt\Sigma = \frac{1}{2}\frac{\partial}{\partial\t E}\scalar{\gt\str:\t c:\gt\str},\quad
\t M = \frac{1}{2}\frac{\partial}{\partial\gt\chi}\scalar{\gt\str:\t c:\gt\str}.
\]
That is: \comment{the image of $\t A$ is spanned by the resultants $(\gt\Sigma,\t M)$ of membrane stresses $\gt\sigma\equiv\t c:\gt\str$}. A full rigorous proof of the identities in~\eqref{eq:imker} is detailed in the supplemental material.

\subsection*{Microstructure-independent relationships for effective membrane moduli}
The static-geometric isomorphism endows the space of effective isometric deformations with a natural symplectic structure: for any pair of modes $(\t E_i,\gt\chi_i)_{i=1,2}$ in that space
\[\label{eq:symplectic}
    \t E_1:\cof\gt\chi_2 - \gt\chi_1:\cof\t E_2 = 0.
\]
This is a restatement of the Hill-Mandel lemma~\eqref{eq:HM} with the isomorphism~\eqref{eq:iso} in mind and with $\gt\str=\t 0$ given that the involved strains are isometric. Such a relationship was recently derived using purely geometric tools in cases where either mode is a pure membrane mode, e.g., $\gt\chi_1=\t 0$, so that
\[\label{eq:orthogonality}
    \t E_1:\cof\gt\chi_2 = 0;
\]
see~\cite{Nassar2024,Nassar2024philo}. In other words, the space of effective isometric deformations that are pure membrane modes is orthogonal to the space of isometric bending strains (rotated through $90^\circ$). This orthogonality, alone, shows that there can be no more than three effective isometric deformations. Here, with the help of the static-geometric analogy, and the minimum energy formulation of the homogenization problem, existence of exactly three modes is proven.

Equation~\eqref{eq:orthogonality} can be recast as an equality between in-plane and out-of-plane Poisson's coefficients. This has been explored by many authors in many cases; see, e.g.,~\cite{Schenk2013,Wei2013,Nassar2017a,Pratapa2019,Nassar2022,Xu2024}. Thanks to the isomorphism constructed by the analogy, it is possible to draw a whole series of exact relationships among the effective moduli of a periodic membrane, all summarized into
\[\label{eq:exact}
    \t A \begin{bmatrix}
    \t 0 & -\cof \\ \cof & \t 0
\end{bmatrix} \t A = \t 0.
\]
This relationship is more robust than the counting rule strictly speaking. Indeed, one can reasonably assume that the effective stiffness tensor $\t A$ depends continuously on the local membrane stiffness~$\t c$ which can then be taken to zero at places so as to create holes. By contrast, the above relationship can certainly fail in the presence of handles, in which case $\t A$ can become full rank (e.g., as in foldcore sandwich panels~\cite{Lebee2012} and in Figure~\ref{fig:spectrum}a).

Equation~\eqref{eq:exact} is a ``microstructure-independent'' exact relationship. It is reminiscent of earlier works in the theory of composites~\cite{Milton2002}. More so, several such works employ what is referred to as a ``translation'', i.e., a transformation that maps fields on the right side of a constitutive equation (e.g., $\gt\str$ in $\gt\sigma=\t c:\gt\str$) onto fields on the left side of the same equation (e.g., $\gt\sigma$). In 2D elasticity, a ``translation'' is provided by none other than the $\cof$ operator, i.e., a pointwise $90^\circ$ rotation~\cite{Cherkaev1992}. The same ``translation'' is also familiar to practitioners of the art of graphic statics~\cite{Mcrobie2016,Mitchell2016,Tachi2015}. Therein, the ``translation'' rotates admissible fields of bar tensions into compatible fields of infinitesimal rotations. 

\subsection*{The role of topology}
The topology of the shell plays a crucial role that remained implicit in the presentation above and that should be highlighted. On one hand, suppose the shell had handles, then there would be topological obstructions to integrating an admissible stress $\gt\sigma$ into an Airy stress function $\phi$. The static-geometric analogy, for averages, will then only encapsulate a subspace of admissible states of stress. The rule needs to be weakened into
\[
    \dim \{(\t E_o,\gt\chi_o)\} \leq 3 \quad \text{(with handles, \comment{no holes})}.
\]
On the other hand, suppose the shell had holes, then there would be extra admissibility constraints for stresses in the form of zero tractions at boundaries. The analogy fails by leaving out isometries that do not reproduce this condition, i.e., isometries that cannot be extended over the holes or, equivalently, that deform the holes. The rule then weakens into
\[
     \dim \{(\t E_o,\gt\chi_o)\} \geq 3 \quad \text{(with holes, \comment{no handles})}.
\]

\begin{figure}
    \centering
    \includegraphics[width=\linewidth]{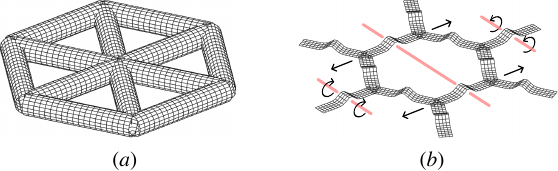}
    \caption{\comment{Influence of topology beyond simple connectivity: (a) example of a surface with handles (three per unit cell) that has no effective isometries; (b) example of a surface with holes (one connected component per unit cell) that has six effective isometries. Both surfaces are a collection of cylinders with three sets of parallel axes (one of which is drawn) but with different topologies. Each family of closed cylinders in (a) can resist one uniaxial membrane stress and one uniaxial bending stress meaning, by the static-geometric analogy, that it admits no effective isometries. Each family of open cylinders in (b) can be isometrically flattened (following straight arrows) or bent (following circular arrows) to accommodate six (two modes, three axes) effective isometries.}}
    \label{fig:spectrum}
\end{figure}

\comment{To illustrate the way in which topology influences the count, consider the example of a ``thickened'' triangular truss as depicted on Figure~\ref{fig:spectrum}a. This surface is not simply connected (with three handles per unit cell) and, quite intuitively, has no effective isometric deformations. At the other extreme, consider the ``cork-inspired'' honeycomb of Figure~\ref{fig:spectrum}b. The wiggles, being cylindrical, allow for uniaxial stretching in three directions orthogonal to their axes as well as for uniaxial bending in three directions about their axes. Thus, this surface admits six effective isometries, i.e., the maximum possible. Clearly, this surface is not simply connected (with one connected boundary component per unit cell). Simply connected surfaces, as shown here, are precisely in the middle of this spectrum with three zero modes.}

\subsection*{The unlikelihood of a conjecture}
\begin{figure*}[ht!]
    \centering
    \includegraphics[width=\linewidth]{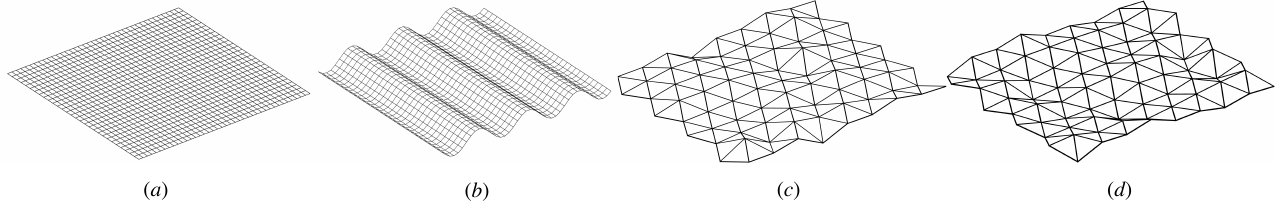}
    \caption{\comment{Surfaces with (a) zero, (b) one, (c) two and (d) three pure membrane modes. Surfaces in (c) and (d) are obtained using the numerical optimization procedure described in the text; their count is not known to be exact albeit strongly suggested by the numerics.}}
    \label{fig:optim}
\end{figure*}

\comment{The three effective isometries of a simply connected surface generally have both membrane ($\t E_o\neq \t 0$) and bending ($\gt\chi_o\neq \t 0$) components. If not, then we say that they are a \emph{pure membrane} (if $\gt\chi_o=\t 0$) and a \emph{pure flexure} (if $\t E_o=\t 0$). The plane is an example of a surface that has no pure membrane modes and three pure flexure modes. The simple corrugation is an example with a single pure membrane mode and two pure flexure modes; see Figure~\ref{fig:optim}a,~b. Beyond that, there are no trivial examples with two or three pure membrane modes.} It is not known if there is a fundamental limit on the number of pure membrane modes, but this appears to be unlikely. Here, the unit cells of surfaces with two and three \emph{approximately} pure membrane modes are produced by a numerical minimization procedure. \comment{To do so, the surface is discretized into a truss and the $6\times 6$ effective elasticity matrix $\t A$ is computed under periodic boundary conditions. The effective moduli corresponding to the target effective isometries are then weighed against the other moduli. The ratio, called $r$, provides an objective function that is minimized by a standard routine. It is given by
\[
r = \frac{\norm{[A_{ij}]_{i\in I, j\in\{1\dots6\}}}}{\norm{[A_{ij}]_{i,j=\{1\dots 6\}\setminus I}}}.
\]
To obtain two pure membrane modes and a single pure flexure mode (a twist), we set $I=\{1,2,6\}$ corresponding to two uniaxial stretches and one twist. To obtain three pure membrane modes, we set $I=\{1,2,3\}$ corresponding to all membrane modes.
Either way, the optimization parameter is the elevation map above a fixed triangular lattice, centered and normalized to avoid spurious elevations:
\[
    \min r \text{ over } \{z \text{ periodic} \mid \scalar{z}=0, \,\max z = z_M\}.
\]
See~\footnote{Code is available at \url{https://github.com/nassarh/discreteMembranes}} for implementation details. Figure~\ref{fig:optim} shows the results. The optimal geometries achieve a relative reduction in the moduli by a factor of $r\approx 10^{-8}$.}

\comment{The above results strongly suggest that there are no restrictions on what effective isometries are realizable, beyond the implications of the simply connected topology, i.e., the count and the symplectic identity~\eqref{eq:symplectic}. In other words, it is suggested that any 3D subspace of $S^2\times S^2$, whose elements satisfy~\eqref{eq:symplectic}, is realizable as the space of effective isometries of some simply connected surface. In the jargon of symplectic linear algebra, such subspaces are called Lagrangians. Interestingly, Lagrangian subspaces are maximal: if the surface were to gain additional isometries, by weakening its topology, then the gained isometry will not satisfy~\eqref{eq:symplectic}. The set of all Lagrangian subspaces, i.e., of all candidate spaces of effective isometries, is the Lagrangian Grassmannian and is known to be a manifold of dimension six (in the present case). This means that the subspace of effective isometries depends continuously on a set of six coordinates. These coordinates are the $3\times 6$ components of a set of three zero modes $(\t E,\gt \chi)$, minus the three constraints provided by identity~\eqref{eq:symplectic}, and taken modulo the 9-dimensional general linear group of $\R^3$ corresponding to transformations that shuffle and combine the zero modes but leave the subspace as is. See~\cite{Meinrenken2024} for more details on Lagrangian Grassmannians.}

\subsection*{Extensions}
It is worthwhile to discuss how the results generalize to cases where the shell's midsurface is piecewise smooth, i.e., with jumps in the tangent plane along crease lines (or crease curves). In these cases, admissible stresses and compatible strains can be discontinuous, albeit with precise jumps that guarantee local force balance and displacement continuity. Specifically, \emph{normal} stresses and \emph{tangent} strains remain continuous across crease lines. The normal and tangent directions being at $90^\circ$ to one another, the static-geometric analogy remains valid. The Hill-Mandel lemma remains valid as well, in the same fashion that it is valid in the presence of perfect material interfaces. See~\cite{Nassar2024philo,Bouchitt2019,Amendola2023} for more detail. By the analogy and the lemma, the count holds. For non-periodic, statistically homogeneous, shells such as wrinkled or rough thin material surfaces, the Hill-Mandel lemma is known to remain valid in the limit of fast fluctuations, i.e., when the size of the shell is infinite relative to the representative volume element. Mathematical literature refers to this result as the $\mathrm{div}$-$\mathrm{curl}$ lemma; see~\cite{Fehrman2020}. Passing to the limit of fast fluctuations relieves the lemma, and therefore the result, of the burden of periodic boundary conditions.

\subsection*{Conclusion}
The results presented here establish a counting rule for effective isometric deformations in simply connected shells. By combining the static–geometric analogy with the Hill–Mandel lemma, the spaces of effective isometries and of admissible effective stresses were proven to be isomorphic, orthogonal and Lagrangian (relative to a symplectic form). Their dimensionality then yields precisely three effective isometric modes, just as it yields precisely three admissible effective stresses. The same arguments reveal an exact algebraic relationship between the effective elasticity moduli providing a microstructure-independent constraint on the homogenized constitutive law. Note that the fact that the shell's surface was taken to be a smooth graph is a matter of convenience. By contrast, simple connectedness is an essential, both sufficient and necessary, ingredient of the theory. Indeed, topology is known to severely impact structural stiffness, as it does for instance in Saint-Venant's problem of torsion for simply or multiply connected sections; the present context is no exception.

\bibliography{lib}
\begin{acknowledgments}
Work supported by the NSF under CAREER award No. CMMI-2045881. Part of this work was inspired by a series of talks given at two symposia: Geometry of Materials at ICERM (Brown University) and GASM at IMAG (Universidad de Granada); the author thanks the speakers and organizers.
\end{acknowledgments}

\appendix

\comment{%
\begin{widetext}
\subsection*{Supplemental Material: Proof of the Hill-Mandel lemma~\eqref{eq:HM}}
Consider a field of membrane stresses $\gt\sigma$ that are statically admissible in the sense of~\eqref{eq:eq}. Let $\gt\str$ be the field of membrane strains of an ansatz $(\t u,w)$ satisfying~\eqref{eq:uw}, i.e.,
\[
\begin{bmatrix}
    u_\alpha \\ w
\end{bmatrix} =
\begin{bmatrix}
    E_{\alpha\beta}x_\beta + z\chi_{\alpha\beta}x_\beta \\ -\frac{1}{2}\chi_{\mu\nu}x_\mu x_\nu
\end{bmatrix}
+ \begin{bmatrix}
    \tilde u_\alpha \\ \tilde w
\end{bmatrix}
\]
where the periodic correction has been denoted with tilde. Thus, the membrane strain is
\[
    \gt\str = \t E + z\gt\chi +\gt\nabla^s \tilde{\t u} + \gt\nabla\tilde w\tp^s\gt\nabla z.
\]
Now expand
\begin{align}
     &\scalar{\gt\sigma:\gt\str} - \scalar{\gt\sigma}:\t E - \scalar{z\gt\sigma}:\gt\chi  \\
    =&\scalar{\sigma_{\alpha\beta}(\tilde u_{\alpha,\beta}+\tilde w_\beta z_\alpha)}\tag{by symmetry of $\gt\sigma$}, \\=&\scalar{(\sigma_{\alpha\beta}(\tilde u_{\alpha}+\tilde w z_\alpha))_\beta}-
    \scalar{\sigma_{\alpha\beta,\beta}(\tilde u_{\alpha}+\tilde w z_\alpha)+\sigma_{\alpha\beta} z_{\alpha\beta}\tilde w}\tag{by the product rule}\\
    =& \,0, \tag{by periodicity \& static admissibility}
\end{align}
to get the desired lemma. It is crucial to note that the lemma holds for virtual fields, i.e., regardless of whether $\gt\sigma=\t c:\gt\str$ or not.
\end{widetext}}
\subsection*{Supplemental Material: Proof of equation~\eqref{eq:imker}}
\comment{Begin by proving the polarization identity
\[
\begin{bmatrix}
    \t E_1 \\ \gt\chi_1
\end{bmatrix}^T\t A\begin{bmatrix}
    \t E_2 \\ \gt\chi_2
\end{bmatrix} = 
\scalar{\gt\str_1:\t c:\gt\str_2},
\]
for any $(\t E_i,\gt\chi_i)$ where $\gt\str_i$ minimizes the corresponding average strain energy density. Indeed, for any such strains, $\gt\str_3\equiv\gt\str_1+\gt\str_2$ minimizes the strain energy density for prescribed $(\t E_3\equiv\t E_1+\t E_2,\gt\chi_3\equiv\gt\chi_1+\gt\chi_2)$ by linearity. Thus,
\[
\begin{bmatrix}
    \t E_i \\ \gt\chi_i
\end{bmatrix}^T\t A\begin{bmatrix}
    \t E_i \\ \gt\chi_i
\end{bmatrix} = 
\scalar{\gt\str_i:\t c:\gt\str_i},\quad (i=1,2,3).
\]
Then, expand
\[
\scalar{\gt\str_3:\t c:\gt\str_3}-\scalar{\gt\str_1:\t c:\gt\str_1} - \scalar{\gt\str_2:\t c:\gt\str_2}
\]
and simplify thanks to the symmetry of $\t c$ and $\t A$ to get the polarization identity, as desired.}

Now to prove~\eqref{eq:imker}, proceed by proving a double inclusion for each equality, starting with strains.

\begin{enumerate}
    \item 
Let $(\t E_o,\gt\chi_o)\in\text{Ker}\t A$ and let $\gt\str_o$ be the corresponding membrane strain. Then, by definition $\scalar{\gt\str_o:\t c:\gt\str_o}=0$ and $\gt\str_o=\t 0$ by the positive definiteness of the membrane stiffness $\t c$. Hence, $(\t E_o,\gt\chi_o)$ is the effective strain of an admissible isometric deflection.

\item
Let $(\t E_o,\gt\chi_o)$ be the effective strain of an admissible isometric deflection, i.e., with $\gt\str_o=\t 0$. Then, by the polarization identity
\[
\begin{bmatrix}
        \t E \\ \gt\chi
    \end{bmatrix}^T \t A \begin{bmatrix}
        \t E_o \\ \gt\chi_o
    \end{bmatrix} = \scalar{\gt\str:\t c:\gt\str_o} = 0,
\]
for all $(\t E, \gt\chi)$ with corresponding $\gt\str$. Thus,
\[
\t A \begin{bmatrix}
        \t E_o \\ \gt\chi_o
    \end{bmatrix} = 0,\quad \text{and}\quad (\t E_o,\gt\chi_o)\in\text{Ker}\t A.
\]

\item
\comment{Let $(\gt\Sigma,\t M)\in\text{Im}\t A$. Then there exists $(\t E,\gt\chi)$ such that
\[
    \begin{bmatrix}
        \gt\Sigma \\ \t M
    \end{bmatrix} = 
    \t A \begin{bmatrix}
        \t E \\ \gt\chi
    \end{bmatrix}.
\]
Let $\gt\sigma=\t c:\gt\str$ be the membrane stresses obtained by the minimization of strain energy~\eqref{eq:min}. Similarly, let $\hat{\gt\str}$ be optimal for a test strain $(\hat{\t E},\hat{\gt\chi})$. The polarization identity yields the first equality in
\[
    \begin{bmatrix}
        \hat{\t E} \\ \hat{\gt\chi}
    \end{bmatrix}^T 
    \begin{bmatrix}
        \gt \Sigma \\ \t M
    \end{bmatrix} =
    \scalar{\hat{\gt\str}:\gt\sigma} = \hat{\t E}:\scalar{\gt\sigma} + \hat{\gt\chi}:\scalar{z\gt\sigma}
\]
whereas the second is due to the Hill-Mandel lemma. As $(\hat{\t E},\hat{\gt\chi})$ is arbitrary, it comes that
\[
\gt\Sigma = \scalar{\gt\sigma},\quad \t M = \scalar{z\gt\sigma}.
\]
Hence, any $(\gt\Sigma,\t M)$ in the image of $\t A$ is an admissible effective state of stress.}
\item
Let $(\gt\Sigma,\t M)$ be an admissible effective stress obtained by averaging an admissible periodic membrane stress $\gt\sigma$. Then, by the Hill-Mandel lemma, 
\[
    \gt\Sigma:\t E_o + \t M:\gt\chi_o = 0,
\]
for all $(\t E_o,\gt\chi_o)\in\text{Ker}\t A$.
Then, $(\gt\Sigma,\t M)\in(\text{Ker}\t A)^\perp=\text{Im}\t A$, by symmetry of $\t A$.
\end{enumerate}
\comment{This concludes the proof that
\[
\begin{split}
\text{Ker} \t A &= \{(\scalar{z\t H_w},-\scalar{\t H_w}) \mid w \text{ is isometric}\},\\
\text{Im} \t A &= \{(\scalar{\gt\sigma},-\scalar{z\gt\sigma})\mid \gt\sigma \text{ is admissible}\}.
\end{split}
\]}
\end{document}